\documentclass[pra,preprint]{revtex4}

\usepackage{graphicx}
\newcommand{\bei}{\begin{itemize}}
\newcommand{\eei}{\end{itemize}}
\newcommand{\bee}{\begin{enumerate}}
\newcommand{\eee}{\end{enumerate}}

\begin{document}

\title{The Generation and Exchange of Entanglement via Zeeman Effect}
\author{Jian-da Wu$^1$, Jian-lan Chen$^1$, Yong-de Zhang$^{1,2}$ }

\affiliation{$^1$Department of Modern Physics, University of Science
and Technology of China, Hefei 230026, People¡¯s Republic of China
\\$^2$CCAST (World Laboratory), P.O.Box 8730, Beijing 100080,
People's Republic of China}

\date{\today}

\begin{abstract}
In this paper we show a new way to generate entanglement via two
identical three-level atoms splitting in the magnetic field
interacting with the cavity field. By the system we investigate, We
can acquire the EPR state, multi-dimensional entangled states
$\emph{etc.}$ which are more stable than usual realization by
high-energy-level Rydberg atoms and we can realize the local
exchange operator too. We also achieve the goal of maintaining long-
time entanglement between atoms. At last, by using the procedure of
local exchange, we put forward an experimental scheme for quantum
feedback.
\end{abstract}

\pacs{42.50.Pq; 03.67.-a; 03.67.Mn; 32.60.+i; 42.50.-p}

\maketitle
%---------------------------------------------------------------------------------------

\section{introduction} It is well known that entanglement has been
variously studied and how to get entanglement fast and stably plays
key roles in the quantum information processing.
Multipartite-entanglement is a great resource which is not only of
importance for test of quantum mechanics against local hidden
theory[1], but also useful in quantum teleportation, dense coding
and quantum cryptography[2]. Most of research in quantum
non-locality and quantum information is based on entanglement of
two-level particles. Entangled states for two-level particles have
been observed for photons, atoms in cavity QED, and ions in
trap[3,4]. Zou et al.[5]have present a scheme to generate a
maximally entangled state of two three-level atoms with a
non-resonant cavity by cavity collisions. A scheme for generating
entangled states for multilevel atoms in a thermal cavity has been
proposed[6]. Teleporting entanglement of cavity-field states has
also been presented[7]. Recently, the system of three-level atom in
two-mode field has been exactly calculated[8]. Most of the work
concentrate on two-level atoms' manipulation and evolution with the
cavity field. Although a scheme for generating entangled states for
multilevel atoms has been proposed[6], its content still bases on
the two-level atoms.

In this paper, we give a new way to realize the entanglement of
three-level atoms, we investigate two ``L" type three-level atoms
whose spin are both zero interacting with the cavity field in the
magnetic field. First, we consider two identical three-level atoms
Zeeman-splitting in the magnetic field plus the cavity field, As it
will be shown, we can easily acquire EPR states, multi-dimensional
entangled states, $\emph{etc}$.The atoms we use here are normal
atoms which are different from three-level Rydberg atoms with high
energy levels. It will be more convenient to realize the
entanglement between atoms and the entanglement is more stable. And
at the end of our paper, we present an experimental scheme for
quantum feedback.

\section{generation of entanglement}
We exploit two zero-spin atoms interacting with electromagnetic
field in a stable-constant magnetic field. In each atom one electron
has been excited to $\emph p$ state. For convenience we denote
$|ij\rangle$ as the state with $l=i,m=j$. Due to electric dipole
transition rule, the transition in the subspace of $l=1$ will not be
allowed. The transition can be allowed only between the subspaces of
$l=1$ and $l=0$. The Lamb shift of the atoms can be ignored because
it is far smaller than the system's resonant energy. Under this
approximation, only subsequent transitions $|1,\pm
1\rangle\leftrightarrow|0 0\rangle$ will appear. Then the
interaction between the atoms and the electromagnetic field can be
written as
\begin{equation}
g\left[ {\left( {l_1^ +   + l_2^ +  } \right)a + \left( {l_1^ - +
l_2^ - } \right)a^\dag  } \right]   \label{g1}
\end{equation}
where g is the interaction coefficient between the atoms and the
electromagnetic field, and $l_i^ +   = | 1 \rangle _i\langle 0 | + |
0 \rangle_i \langle { - 1} |$, $l_i^ -   = \left( {l_i^ +  }
\right)^\dag$

The interaction between the atoms and the magnetic field and the
interaction between the atoms are $- \left(
{\mathord{\buildrel{\lower3pt\hbox{$\scriptscriptstyle\rightharpoonup$}}
\over \mu } _1  +
\mathord{\buildrel{\lower3pt\hbox{$\scriptscriptstyle\rightharpoonup$}}
\over \mu } _2 } \right) \cdot
\mathord{\buildrel{\lower3pt\hbox{$\scriptscriptstyle\rightharpoonup$}}
\over B}$,$\quad\lambda
\mathord{\buildrel{\lower3pt\hbox{$\scriptscriptstyle\rightharpoonup$}}
\over \mu } _1  \cdot
\mathord{\buildrel{\lower3pt\hbox{$\scriptscriptstyle\rightharpoonup$}}
\over \mu } _2$ respectively. Suppose the direction of the magnetic
field is along $z$ axis, we can simplify (3) as below
\begin{equation}
- \left( {\mu _{1z}  + \mu _{2z} } \right)B_z  + \lambda \mu _{1z}
\mu _{2z} \label{g2}
\end{equation}
where $\lambda$ is the interaction coefficient between the magnetic
moments. The equation above can be rewritten as
\begin{equation}
\beta (l_{1z}  + l_{2z} ) + \alpha l_{1z}  \cdot l_{2z} \label{g3}
\end{equation}
where $\beta  = e/{2\mu }$ and $\alpha  = \lambda e^2 /{4\mu ^2 }$.
And finally we obtain the Hamiltonian of the system
\begin{eqnarray}
H = &\omega& a^\dag  a + \beta \left( {l_{1z}  + l_{2z} } \right)
\\ \nonumber
 &+&
g\left[ {\left( {l_1^ +   + l_2^ +  } \right)a + \left( {l_1^ - +
l_2^ -  } \right)a^\dag  } \right] + \alpha l_{1z}l_{2z}  \label{g4}
\end{eqnarray}
$\omega$ is the frequency of the electromagnetic field.

The Hamiltonian in the interaction picture is
\begin{eqnarray}
H_i^{(I)}=&\alpha& l_{1z} l_{2z}+\\
\nonumber &g&\left[{e^{i\left( {\beta  - \omega } \right)t} \left(
{l_1^ + + l_2^ + } \right)a + e^{ - i\left( {\beta  - \omega }
\right)t} \left( {l_1^ - + l_2^ -  } \right)a^\dag  } \right]
\label{g5}
\end{eqnarray}
Then consider the exact resonance, we can neglect the phase terms
${e^{i\left( {\beta  - \omega } \right)t} }$ and ${e^{-i\left(
{\beta - \omega } \right)t} }$. The interaction Hamiltonian in the
interaction picture can be simplified to
\begin{eqnarray}
H_i^{(I)}  = g\left[ {\left( {l_1^ +   + l_2^ +  } \right)a + \left(
{l_1^ -   + l_2^ -  } \right)a^\dag  } \right] + \alpha l_{1z}
l_{2z}   \label{g6}
\end{eqnarray}

It is easy to find that $a^\dag  a + l_{1z}  + l_{2z}$ is the
conversation quantity of the system, thus we can choose states
$\{|N\rangle_p|l_{1z}l_{2z}\rangle_a\}$ as the complete basis of the
system with $a$ and $p$ indicating atoms and photon respectively and
$N$ is the number of  photons. Suppose $a^\dag a + l_{1z} +
l_{2z}=n$ then we can obtain the invariant space of the system. The
complete basis of invariant space are: $| n + 2\rangle_p| - 1, - 1
\rangle_a$, $| n + 1\rangle_p|0, - 1 \rangle_a$, $| n + 1\rangle_p|
- 1,0\rangle_a $, $| n\rangle_p| 1, - 1\rangle_a$, $| n\rangle_p|
0,0 \rangle_a$, $| n\rangle_p|  - 1,1 \rangle_a $, $| n -
1\rangle_p| 1,0 \rangle_a$, $| n - 1\rangle_p| 0,1 \rangle_a$, $| n
- 2\rangle_p| 1,1 \rangle_a$. Thus, in this invariant subspace, the
interaction Hamiltonian in the interaction picture can be expressed
as $9 \times 9$ matrix below

\[
\left( {\begin{array}{ccccccccc}
   \alpha & {g\sqrt {n + 2} } & {g\sqrt {n + 2} } & 0 & 0 & 0 & 0 & 0 & 0  \\
   {g\sqrt {n + 2} } & 0 & 0 & {g\sqrt {n + 1} } & {g\sqrt {n + 1} } & 0 & 0 & 0 & 0  \\
   {g\sqrt {n + 2} } & 0 & 0 & 0 & {g\sqrt {n + 1} } & {g\sqrt {n + 1} } & 0 & 0 & 0  \\
   0 & {g\sqrt {n + 1} } & 0 & { - \alpha } & 0 & 0 & {g\sqrt n } & 0 & 0  \\
   0 & {g\sqrt {n + 1} } & {g\sqrt {n + 1} } & 0 & 0 & 0 & {g\sqrt n } & {g\sqrt n } & 0  \\
   0 & 0 & {g\sqrt {n + 1} } & 0 & 0 & { - \alpha } & 0 & {g\sqrt n } & 0  \\
   0 & 0 & 0 & {g\sqrt n } & {g\sqrt n } & 0 & 0 & 0 & {g\sqrt {n - 1} }  \\
   0 & 0 & 0 & 0 & {g\sqrt n } & {g\sqrt n } & 0 & 0 & {g\sqrt {n - 1} }  \\
   0 & 0 & 0 & 0 & 0 & 0 & {g\sqrt {n - 1} } & {g\sqrt {n - 1} } & \alpha   \\
\end{array}} \right)
\]
Because $\alpha  \ll \beta$,g, under the fist order approximation,
we can simplify the above matrix with $\alpha=0$. With this
simplification the matrix is still too large, we can choose
appropriate $n$ to simplify the above matrix furthermore,  for
convenience, we can set $n$ to be zero, then the matrix can be
simplified below
\begin{eqnarray}
H_i^{(I)}  = \left( {\begin{array}{*{20}c}
   0 & {g\sqrt 2 } & {g\sqrt 2 } & 0 & 0 & 0  \\
   {g\sqrt 2 } & 0 & 0 & g & g & 0  \\
   {g\sqrt 2 } & 0 & 0 & 0 & g & g  \\
   0 & g & 0 & 0 & 0 & 0  \\
   0 & g & g & 0 & 0 & 0  \\
   0 & 0 & g & 0 & 0 & 0  \\
\end{array}} \right)  \label{g7}
\end{eqnarray}
 Then we can obtain the evolving state in Schrodinger picture:

\begin{eqnarray}
\begin{array}{l}
 \left| {\psi (t)} \right\rangle  = \left. {\exp \left( { - iH_0 t} \right)\exp \left( { - iH_i^{(I)} t} \right)\left| {\psi (0)} \right\rangle } \right|_{n = 0,\omega  = \beta }  \\
 \;\;\;\;\;\;\;\;\; = \left. {\exp \left( { - iH_i^{(I)} t} \right)\left| {\psi (0)} \right\rangle } \right|_{n = 0,\omega  = \beta }  \\
 \;\;\;\;\;\;\;\;\; =  \\
 \end{array}    \nonumber
\end{eqnarray}

\begin{eqnarray}
\left( {\begin{array}{*{20}c}
   {\frac{1}{7}\left[ {3 + 4\cos \left( {\sqrt 7 gt} \right)} \right]} & { - i\sqrt {\frac{2}{7}} \sin \left( {\sqrt 7 gt} \right)} & { - i\sqrt {\frac{2}{7}} \sin \left( {\sqrt 7 gt} \right)}  \\
   { - i\sqrt {\frac{2}{7}} \sin \left( {\sqrt 7 gt} \right)} & {\frac{1}{2}\left[ {\cos \left( {gt} \right) + \cos \left( {\sqrt 7 gt} \right)} \right]} & {\frac{1}{2}\left[ { - \cos \left( {gt} \right) + \cos \left( {\sqrt 7 gt} \right)} \right]}  \\
   { - i\sqrt {\frac{2}{7}} \sin \left( {\sqrt 7 gt} \right)} & {\frac{1}{2}\left[ { - \cos \left( {gt} \right) + \cos \left( {\sqrt 7 gt} \right)} \right]} & {\frac{1}{2}\left[ {\cos \left( {gt} \right) + \cos \left( {\sqrt 7 gt} \right)} \right]}  \\
   {\frac{{\sqrt 2 }}{7}\left[ { - 1 + \cos \left( {\sqrt 7 gt} \right)} \right]} & { - \frac{{i\sqrt 7 }}{{14}}\left[ {\sqrt 7 \sin \left( {gt} \right) + \sin \left( {\sqrt 7 gt} \right)} \right]} & {\frac{{i\sqrt 7 }}{{14}}\left[ {\sqrt 7 \sin \left( {gt} \right) - \sin \left( {\sqrt 7 gt} \right)} \right]}  \\
   {\frac{{2\sqrt 2 }}{7}\left[ { - 1 + \cos \left( {\sqrt 7 gt} \right)} \right]} & { - i\frac{{\sin \left( {\sqrt 7 gt} \right)}}{{\sqrt 7 }}} & { - i\frac{{\sin \left( {\sqrt 7 gt} \right)}}{{\sqrt 7 }}}  \\
   {\frac{{\sqrt 2 }}{7}\left[ { - 1 + \cos \left( {\sqrt 7 gt} \right)} \right]} & {\frac{{i\sqrt 7 }}{{14}}\left[ {\sqrt 7 \sin \left( {gt} \right) - \sin \left( {\sqrt 7 gt} \right)} \right]} & { - \frac{{i\sqrt 7 }}{{14}}\left[ {\sqrt 7 \sin \left( {gt} \right) + \sin \left( {\sqrt 7 gt} \right)} \right]}  \\
\end{array}} \right.  \nonumber
\end{eqnarray}
\begin{eqnarray}
\left. {\begin{array}{*{20}c}
   {\frac{{\sqrt 2 }}{7}\left[ { - 1 + \cos \left( {\sqrt 7 gt} \right)} \right]} & {\frac{{2\sqrt 2 }}{7}\left[ { - 1 + \cos \left( {\sqrt 7 gt} \right)} \right]} & {\frac{{\sqrt 2 }}{7}\left[ { - 1 + \cos \left( {\sqrt 7 gt} \right)} \right]}  \\
   { - \frac{{i\sqrt 7 }}{{14}}\left[ {\sqrt 7 \sin \left( {gt} \right) + \sin \left( {\sqrt 7 gt} \right)} \right]} & { - i\frac{{\sin \left( {\sqrt 7 gt} \right)}}{{\sqrt 7 }}} & {\frac{{i\sqrt 7 }}{{14}}\left[ {\sqrt 7 \sin \left( {gt} \right) - \sin \left( {\sqrt 7 gt} \right)} \right]}  \\
   {\frac{{i\sqrt 7 }}{{14}}\left[ {\sqrt 7 \sin \left( {gt} \right) - \sin \left( {\sqrt 7 gt} \right)} \right]} & { - i\frac{{\sin \left( {\sqrt 7 gt} \right)}}{{\sqrt 7 }}} & { - \frac{{i\sqrt 7 }}{{14}}\left[ {\sqrt 7 \sin \left( {gt} \right) + \sin \left( {\sqrt 7 gt} \right)} \right]}  \\
   {\frac{1}{{14}}\left[ {6 + 7\cos \left( {gt} \right) + \cos \left( {\sqrt 7 gt} \right)} \right]} & {\frac{1}{7}\left[ { - 1 + \cos \left( {\sqrt 7 gt} \right)} \right]} & {\frac{1}{{14}}\left[ {6 - 7\cos \left( {gt} \right) + \cos \left( {\sqrt 7 gt} \right)} \right]}  \\
   {\frac{1}{7}\left[ { - 1 + \cos \left( {\sqrt 7 gt} \right)} \right]} & {\frac{1}{7}\left[ {5 + 2\cos \left( {\sqrt 7 gt} \right)} \right]} & {\frac{1}{7}\left[ { - 1 + \cos \left( {\sqrt 7 gt} \right)} \right]}  \\
   {\frac{1}{{14}}\left[ {6 - 7\cos \left( {gt} \right) + \cos \left( {\sqrt 7 gt} \right)} \right]} & {\frac{1}{7}\left[ { - 1 + \cos \left( {\sqrt 7 gt} \right)} \right]} & {\frac{1}{{14}}\left[ {6 + 7\cos \left( {gt} \right) + \cos \left( {\sqrt 7 gt} \right)} \right]}  \\
\end{array}} \right){\left| {\psi (0)} \right\rangle } \nonumber\\
\label{g8}
\end{eqnarray}
where $|\psi(0)\rangle$ is initial state with $n = 0,\omega  =
\beta$.

We can choose proper initial state to decide the number of $n$. Next
we show how to get entanglement sources via zeeman effect. We choose
$|0\rangle_p|1,-1 \rangle_a$ as the initial state of the system and
the state we choose here is also easy to be implemented in actual
experiment, where the first zero is the photon-number state, the
second and the third number in the state indicate the atoms' initial
states. Based on the basis we choose
\begin{eqnarray}
\left| {\psi (0)} \right\rangle  = |0\rangle_p|1,-1 \rangle_a  =
\left( {\begin{array}{*{20}c}
   0  \\
   0  \\
   0  \\
   1  \\
   0  \\
   0  \\
\end{array}} \right)\label{g9}
\end{eqnarray}
then we can obtain the evolving state below
\begin{eqnarray}
\begin{array}{l}
 \left| {\psi (t)} \right\rangle  = \frac{{\sqrt 2 }}{7}\left[ { - 1 + \cos \left( {\sqrt 7 gt} \right)} \right]\left| 2 \right\rangle _{p} \left| { - 1, - 1} \right\rangle _{a}  - \frac{{i\sqrt 7 }}{{14}}\left[ {\sqrt 7 \sin \left( {gt} \right) + \sin \left( {\sqrt 7 gt} \right)} \right]\left| 1 \right\rangle _{p} \left| {0, - 1} \right\rangle _{a}  +  \\
 \frac{{i\sqrt 7 }}{{14}}\left[ {\sqrt 7 \sin \left( {gt} \right) - \sin \left( {\sqrt 7 gt} \right)} \right]\left| 1 \right\rangle _{p} \left| { - 1,0} \right\rangle _{a}  + \frac{1}{{14}}\left[ {6 + 7\cos \left( {gt} \right) + \cos \left( {\sqrt 7 gt} \right)} \right]\left| 0 \right\rangle _{p} \left| {1, - 1} \right\rangle _{a}  +  \\
 \frac{1}{7}\left[ { - 1 + \cos \left( {\sqrt 7 gt} \right)} \right]\left| 0 \right\rangle _{p} \left| {0,0} \right\rangle _{a}  + \frac{1}{{14}}\left[ {6 - 7\cos \left( {gt} \right) + \cos \left( {\sqrt 7 gt} \right)} \right]\left| 0 \right\rangle _{p} \left| { - 1,1} \right\rangle _{a}  \\
  \\
 \end{array}  \label{g10}
 \end{eqnarray}
When $t =\frac{{2n\pi }}{{\sqrt 7 g}}$, the state evolves to
\begin{eqnarray}
\begin{array}{l}
 \left| {\psi (t)} \right\rangle  =  - \frac{i}{2}\left[ {\sin \left( {\frac{{2n\pi }}{{\sqrt 7 }}} \right)} \right]\left| 1 \right\rangle _{p} \left| {0, - 1} \right\rangle _{a}  + \frac{i}{2}\left[ {\sin \left( {\frac{{2n\pi }}{{\sqrt 7 }}} \right)} \right]\left| 1 \right\rangle _{p} \left| { - 1,0} \right\rangle _{a}  \\
  + \frac{1}{2}\left[ {1 + \cos \left( {\frac{{2n\pi }}{{\sqrt 7 }}} \right)} \right]\left| 0 \right\rangle _{p} \left| {1, - 1} \right\rangle _{a}  + \frac{1}{2}\left[ {1 - \cos \left( {\frac{{2n\pi }}{{\sqrt 7 }}} \right)} \right]\left| 0 \right\rangle _{p} \left| { - 1,1} \right\rangle _{a}  \\
 \end{array}\label{g11}
\end{eqnarray}
by detecting the photons, we can obtain the two-atom entangled state
immediately
\begin{eqnarray}
\frac{1}{{\sqrt 2 }}\left( {\left| {0, - 1} \right\rangle _{12}  -
\left| { - 1,0} \right\rangle _{12} } \right)  \label{g12}
\end{eqnarray}
Now we change the direction of the electromagnetic field and let it
along x axis. After setting appropriate time we can obtain the
entangled state
\begin{equation}
\frac{1}{\sqrt{2}}(|1,-1\rangle_{12}+|-1,1\rangle_{12})\label{g13}
\end{equation}
With the same process we can obtain the  entangled state$
\frac{1}{\sqrt{2}}(|1,1\rangle_{12}+|-1,1\rangle_{12})$ with the
electromagnetic field along y axis.

\section{Realization of local exchange operator}
 Furthermore, using the system we study, we can
easily realize long-time entanglement between atoms and realize a
local exchange. By the step below we can keep the information in the
entangled atoms. Suppose the initial entangled state is $c_1 \left|
{0, - 1} \right\rangle _{12} + c_2 \left| { - 1,0} \right\rangle
_{12}$ and these two atoms are in two cavity, respectively. Before
the spontaneous radiation, we send two atoms whose states are both
in $\left| { - 1} \right\rangle$ into the two cavity respectively.
In each cavity the conversation number $n=-1$, as it is shown above,
we can easily get the evolving state as below
\begin{eqnarray}
\left| {\psi (t)} \right\rangle  = e^{i\omega t} \left(
{\begin{array}{*{20}c}
   {\cos \left( {\sqrt 2 gt} \right)} & { - \frac{{i\sin \left( {\sqrt 2 gt} \right)}}{{\sqrt 2 }}} & { - \frac{{i\sin \left( {\sqrt 2 gt} \right)}}{{\sqrt 2 }}}  \\
   { - \frac{{i\sin \left( {\sqrt 2 gt} \right)}}{{\sqrt 2 }}} & {\frac{1}{2}\left[ {1 + \cos \left( {\sqrt 2 gt} \right)} \right]} & {\frac{1}{2}\left[ { - 1 + \cos \left( {\sqrt 2 gt} \right)} \right]}  \\
   { - \frac{{i\sin \left( {\sqrt 2 gt} \right)}}{{\sqrt 2 }}} & {\frac{1}{2}\left[ { - 1 + \cos \left( {\sqrt 2 gt} \right)} \right]} & {\frac{1}{2}\left[ {1 + \cos \left( {\sqrt 2 gt} \right)} \right]}  \\
\end{array}} \right)\left. {\left| {\psi (0)} \right\rangle } \right|_{n =  - 1,\omega  = \beta
}  \label{g14}
\end{eqnarray}
when the evolving time of the system pass through $t = \frac{{\left(
{2n + 1} \right)\pi }}{{\sqrt 2 g}}$, we can obtain the exchange of
states as below
\[
|\psi (0)\rangle  = | 0\rangle_p|0, - 1\rangle_a= \left(
{\begin{array}{*{20}c}
   0  \\
   1  \\
   0  \\
\end{array}}\right )
\to {| {\psi (t)} \rangle }|_{t = \frac{{\left( {2n + 1} \right)\pi
}}{{\sqrt 2 g}}}
 = | 0\rangle_p| - 1,0 \rangle_a  = \left( {\begin{array}{*{20}c}
   0  \\
   0  \\
   1  \\
\end{array}} \right)
\]

\[
|\psi (0)\rangle  = | 0\rangle_p|- 1,0\rangle_a= \left(
{\begin{array}{*{20}c}
   0  \\
   0  \\
   1  \\
\end{array}}\right )
\to {| {\psi (t)} \rangle }|_{t = \frac{{\left( {2n + 1} \right)\pi
}}{{\sqrt 2 g}}}
 = | 0\rangle_p|0, - 1 \rangle_a  = \left( {\begin{array}{*{20}c}
   0  \\
   1  \\
   0  \\
\end{array}} \right)
\]
As is shown above, we can realize local exchange process. Using this
property we can keep long-time entanglement as below
\begin{eqnarray}
\begin{array}{l}
 \left( {c_1 \left| {0, - 1} \right\rangle _{12}  + c_2 \left| { - 1,0} \right\rangle _{12} } \right)\left| { - 1, - 1} \right\rangle _{34}  = c_1 \left| {0, - 1} \right\rangle _{13} \left| { - 1, - 1} \right\rangle _{24}  + c_2 \left| {0, - 1} \right\rangle _{24} \left| { - 1, - 1} \right\rangle _{13}  \\
 \mathop  \to \limits_{evolution}^{t = \frac{{\left( {2n + 1} \right)\pi }}{{\sqrt 2 g}}} c_1 \left| { - 1,0} \right\rangle _{13} \left| { - 1, - 1} \right\rangle _{24}  + c_2 \left| { - 1,0} \right\rangle _{24} \left| { - 1, - 1} \right\rangle _{13}  = \left( {c_1 \left| {0, - 1} \right\rangle _{34}  + c_2 \left| { - 1,0} \right\rangle _{34} } \right)\left| { - 1, - 1} \right\rangle _{12}  \\
 \end{array}   \label{g15}
\end{eqnarray}
where the first and the third atoms are in a cavity, and the second
and the forth atoms are in the other. Both cavities are in vacuum
states. We can see that after a period of evolution time the
entanglement between the first two atoms is transferred to the last
two atoms by proper arrangement of the atoms' time of flight. After
we choose a proper evolution time which is much smaller than the
average spontaneous radiation time, we can persist the entanglement
and store the information in the entangled state for a long time.

\section{An experimental scheme of quantum feedback}
Furthermore, with the local exchange procedure above we can realize
some kind of quantum feedback shown in the figure below. However,
the conception of quantum feedback here is a little different from
the traditional opinions about quantum feedback presented in paper
[10,11]. The problem is that when we have a system which is designed
for consistently producing the maximally entangled states, how we
can obtain the information about the quality of the entangled states
by our experimental system without changing them such as moving the
instrument or adjusting the magnetic field or changing other
necessary parts. In the experimental scheme shown below, with the
local exchange procedure we can not only check the quality of our
experimental system but also give positive feedback for our system.
\begin{center}
\includegraphics[width=16cm,height=12cm]{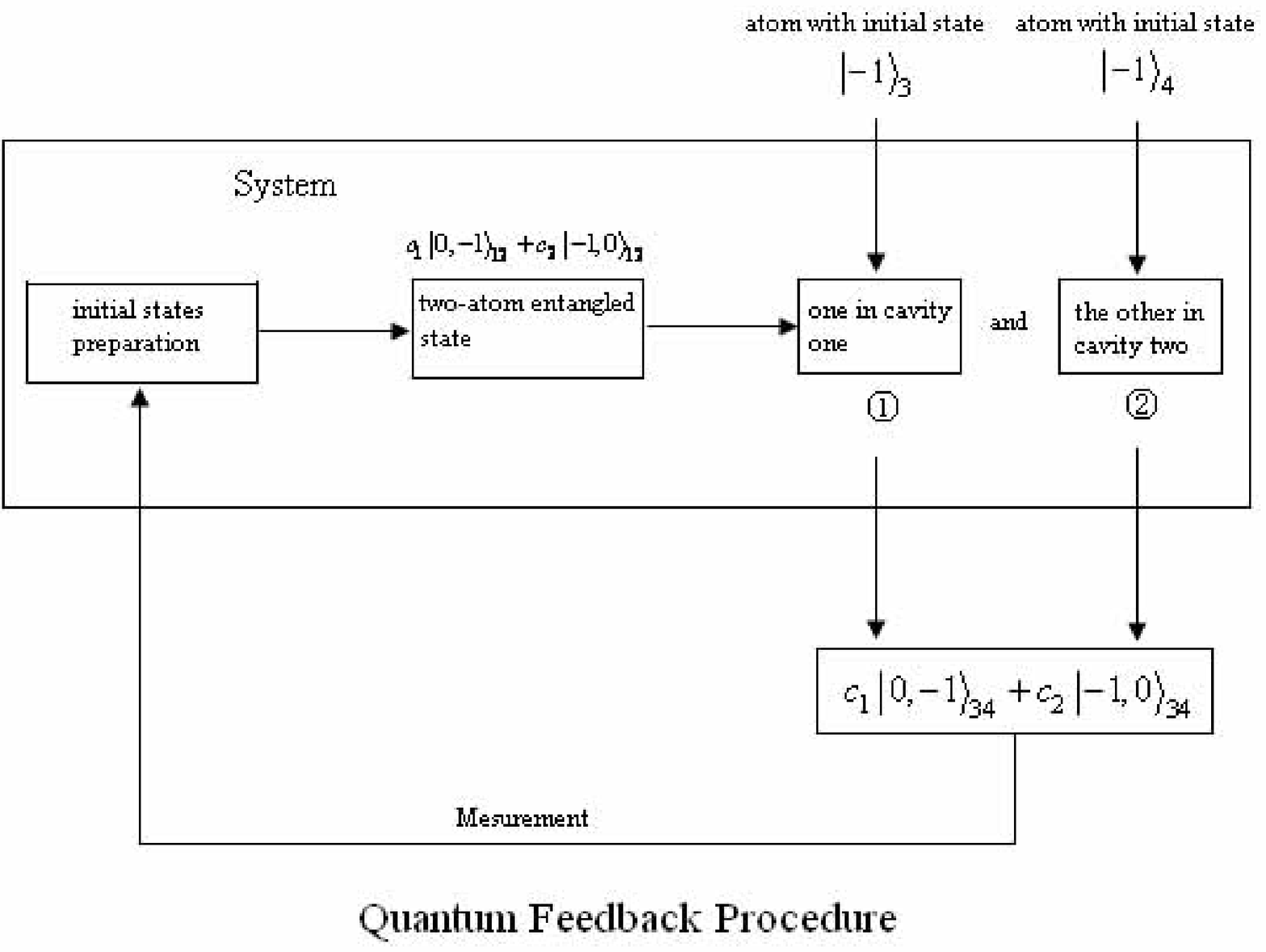}
\end{center}
The generation procedure of the first and the second atoms entangled
state has been shown in sec II. And the procedure that the assisted
third and fourth flight atoms finish the exchange process in the
cavity one and cavity two has been shown in Eq.(\ref{g15}). The
third and fourth atoms finally bring the entangled information of
the first and second atoms out from the experimental system. After
we measure the entangled state of the third and the fourth atoms, we
can acquire the information such as the quality, the stability and
the efficiency of our experimental system without changing them.
From the result of the measurement we can adjust our experimental
system and ameliorate the quality of the productivity of the
experimental system. Such feedback procedure may need some classical
information from the measurement of output states, but it has
advantage that we have no need to "change" or "move" our
experimental system. As we all know that quantum system is so
sensitive and subtle that if we "change" or "move" any part of our
experimental system, the results would be completely different.
Accordingly, following the cyclic feedback procedure above and
accumulating useful improvement, we will finally get a sound system
to produce entangled states.

\section{Conclusions}
In summary, we have proposed a way to realize two-atom maximally
entangled states and generate multi-atoms entangled states. By the
model we use, we realize long-time entanglement and can keep the
information in the entangled state for a long time. Finally, we also
realize local exchange operator. Furthermore, by using the local
exchange procedure, we also present an experimental scheme of
quantum feedback.

\section{acknowledgements}

We thank L. Li for giving help on calculation. Specially thank
Professor S.X. Yu and Doctor M.S. Zhao for illuminating discussions.

%---------------------------------------------------------------------------------------

\end{document}